\documentclass[arXiv]{actapoly}

\usepackage{xcolor}

\begin{document}

\title{On Two Ways to Look for Mutually Unbiased Bases}

\institution{my}{Universit\'e de Lyon, Universit\'e Claude Bernard Lyon 1 et CNRS/IN2P3, Institut de Physique Nucl\'eaire, 4 rue Enrico Fermi, 69622 Villeurbanne, France}

\correspondingauthor{Maurice R. Kibler}{my}{kibler@ipnl.in2p3.fr}

\begin{abstract}
Two equivalent ways of looking for mutually unbiased bases are discussed in this note. The passage 
from the search for $d+1$ mutually unbiased bases in $\mathbb{C}^d$ to the search for $d(d+1)$ vectors 
in $\mathbb{C}^{d^2}$ satisfying constraint relations is clarified. Symmetric informationally complete 
positive-operator-valued measures are briefly discussed in a similar vein. 

\end{abstract}


\maketitle

\section{Introduction}

The concept of mutually unbiased bases (MUBs) plays an important role in finite-dimensional quantum mechanics and quantum information (for more details, see~\cite{Vourdas,Tolar,Kib09,Durt} and references therein). Let us recall that two orthonormal bases $\{ | a \alpha \rangle : \alpha = 0, 1, \ldots, d-1 \}$ and  $\{ | b \beta \rangle : \beta = 0, 1, \ldots, d-1 \}$ in the $d$-dimensional Hilbert space
$\mathbb{C}^d$ (endowed with an inner product denoted as $\langle \; | \; \rangle$) are said to be unbiased if the modulus of the inner product 
$\langle a\alpha | b\beta \rangle$ of any vector $| b\beta \rangle$ with any vector $| a\alpha \rangle$ is equal to $1/{\sqrt{d}}$. It is known that the maximum number of MUBs in $\mathbb{C}^d$ is $d+1$ and that this number is reached when $d$ is a power of a prime integer. In the case where $d$ is not a prime integer, it is not known if one can construct $d+1$ MUBs (see~\cite{Durt} for a review). 

In a recent paper~\cite{Kib13}, it was discussed how the search for $d+1$ mutually unbiased bases in $\mathbb{C}^d$ can be approached via the search for $d(d+1)$ vectors in $\mathbb{C}^{d^2}$ satisfying constraint relations. It is the main aim of this note to make the results in~\cite{Kib13} more precise and to show that the two approaches (looking for $d+1$ MUBs in $\mathbb{C}^d$ or for $d(d+1)$ vectors in $\mathbb{C}^{d^2}$) are entirely equivalent. The central results are presented in Sections 2 and 3. In Section 4, parallel developments for the search of a symmetric informationally complete positive-operator-valued measure (SIC POVM) are considered in the framework of similar approaches. Some concluding remarks are given in the last section.  

\section{The two approaches}

It was shown in~\cite{Kib13} how the problem of finding $d+1$ MUBs in $\mathbb{C}^d$, i.e., $d+1$ bases 
			\begin{eqnarray}
B_a = \{ | a \alpha \rangle : \alpha = 0, 1, \ldots, d-1 \} 
			\label{def Ba1} 
			\end{eqnarray}
satisfying 
			\begin{eqnarray}
| \langle a\alpha | b\beta \rangle | = 
\delta_{\alpha ,\beta }\delta_{a,b} + \frac{1}{\sqrt{d}}(1-\delta_{a,b}) 								
			\label{complete1 def MUBs1} 
			\end{eqnarray}
can be transformed in the problem of finding $d(d+1)$ vectors ${\bf w}(a\alpha)$ in $\mathbb{C}^{d^2}$, of components $w_{pq}(a\alpha)$, satisfying 
			\begin{equation}
\overline{w_{pq}(a\alpha)} = w_{qp}(a\alpha), \quad p,q \in \mathbb{Z}/d\mathbb{Z}
			\label{hermiticity}
			\end{equation} 
			\begin{equation}
\sum_{p = 0}^{d-1}w_{pp}(a\alpha) = 1
			\label{trace}
      \end{equation} 
and
  		\begin{equation}
\sum_{p=0}^{d-1} \sum_{q=0}^{d-1} \overline{w_{pq}(a\alpha)} w_{pq}(b\beta) =
\delta_{\alpha ,\beta }\delta_{a,b} + \frac{1}{d}(1-\delta_{a,b})
			\label{MUBs en w}
  		\end{equation} 
with $a, b = 0, 1, \ldots, d$ and $\alpha, \beta = 0, 1, \ldots, d-1$ in  
(\ref{def Ba1})--(\ref{MUBs en w}). (In this paper, the bar denotes complex conjugation.) This result was described by Proposition 1 in~\cite{Kib13}. In fact, the equivalence of the two approaches (in $\mathbb{C}^d$ and $\mathbb{C}^{d^2}$) requires that each component $w_{pq}(a\alpha)$ be factorized as 
			\begin{eqnarray}
w_{pq}(a\alpha) = \omega_{p}(a\alpha) \overline{\omega_{q}(a\alpha)}
			\label{factorization} 
			\end{eqnarray}
for $a = 0, 1, \ldots, d$ and $\alpha = 0, 1, \ldots, d-1$, a condition satisfied by the example given in~\cite{Kib13}. The factorization of $w_{pq}(a\alpha)$ follows from the fact that the operator $\Pi_{ a\alpha }$ defined in~\cite{Kib13} is a projection operator. 

The introduction of (\ref{factorization}) in (\ref{hermiticity}), (\ref{trace}) and (\ref{MUBs en w}) leads to some simplifications. First, (\ref{factorization}) implies the hermiticity condition (\ref{hermiticity}). Second, by introducing (\ref{factorization}) into (\ref{trace}) and (\ref{MUBs en w}), we obtain 
			\begin{equation}
\sum_{p = 0}^{d-1} \vert \omega_{p}(a\alpha) \vert^2 = 1
			\label{trace en v}
      \end{equation} 
and 
  		\begin{equation}
\left\vert \sum_{p=0}^{d-1} \overline{\omega_{p}(a\alpha)} \omega_{p}(b\beta) \right\vert^2 =
\delta_{\alpha ,\beta }\delta_{a,b} + \frac{1}{d}(1-\delta_{a,b})
			\label{MUBs en v}
  		\end{equation} 
respectively. It is clear that (\ref{trace en v}) follows from (\ref{MUBs en v}) with $a = b$ and $\alpha = \beta$. Therefore, (\ref{hermiticity}) and (\ref{trace en v}) are redundant in view of (\ref{MUBs en w}) and (\ref{factorization}). As a consequence, Proposition 1 in~\cite{Kib13} can be precised and reformulated in the following way. 
 
{\bf Proposition 1.} {\it For $d \geq 2$, finding $d + 1$ MUBs in $\mathbb{C}^d$ (if they exist) is equivalent to finding $d(d + 1)$ vectors ${\bf w} (a\alpha)$ in $\mathbb{C}^{d^2}$, of components $w_{pq}(a\alpha)$ such that
		  \begin{equation}
\sum_{p=0}^{d-1} \sum_{q=0}^{d-1} \overline{w_{pq}(a\alpha)} w_{pq}(b\beta) =
\delta_{\alpha ,\beta }\delta_{a,b} + \frac{1}{d}(1-\delta_{a,b})
			\label{proposition1b}
  		\end{equation} 
and  		
			\begin{equation}
w_{pq}(a\alpha) = \omega_{p}(a\alpha) \overline{\omega_{q}(a\alpha)}, \quad p,q \in \mathbb{Z}/d\mathbb{Z}
			\label{proposition1a}
			\end{equation} 
where $a, b = 0, 1, \ldots, d$ and $\alpha, \beta = 0, 1, \ldots, d-1$.}

This result can be transcribed in matrix form. Therefore, we have the following proposition.

{\bf Proposition 2.} {\it For $d \geq 2$, finding $d + 1$ MUBs in $\mathbb{C}^d$ (if they exist) is equivalent to finding $d(d + 1)$ 
matrices $M_{a\alpha}$ of dimension $d$, with elements 
			\begin{equation}
\left( M_{a \alpha} \right)_{pq} = \omega_{p}(a\alpha) \overline{\omega_{q}(a\alpha)}, \quad p,q \in \mathbb{Z}/d\mathbb{Z}
			\label{proposition2a}
			\end{equation}
and satisfying the trace relations
			\begin{equation}
{\rm Tr} \left( M_{a \alpha} M_{b \beta} \right) = \delta_{\alpha ,\beta}\delta_{a,b} + \frac{1}{d}(1-\delta_{a,b})
			\label{proposition2}
			\end{equation}
where $a, b = 0, 1, \ldots, d$ and $\alpha, \beta = 0, 1, \ldots, d-1$.}

\section{Equivalence}

Suppose that we have a complete set $\{ B_a : a = 0, 1, \ldots, d \}$ of
$d+1$ MUBs in $\mathbb{C}^d$, i.e., $d(d+1)$ vectors $|a \alpha \rangle$ satisfying (\ref{complete1 def MUBs1}), then 
we can find $d(d+1)$ vectors ${\bf w} (a\alpha)$ in $\mathbb{C}^{d^2}$, of components $w_{pq}(a\alpha)$, satisfying (\ref{proposition1b}) and 
(\ref{proposition1a}). This can be achieved by introducing the projection operators 
			\begin{equation}
\Pi _{a\alpha} = | a\alpha \rangle \langle a\alpha |
			\label{Piaalpha}
			\end{equation}
where $a = 0, 1, \ldots, d$ and $\alpha = 0, 1, \ldots, d-1$. In fact, it is sufficient to develop $\Pi _{a\alpha}$ in terms of the 
$E_{pq}$ generators of the $\mathrm{GL}$($d,\mathbb{C}$) complex Lie group; the coefficients of the development are nothing but the 
$w_{pq}(a\alpha)$ complex numbers satisfying (\ref{proposition1b}) and (\ref{proposition1a}), see~\cite{Kib13} for more precisions. 

Reciprocally, should we find $d(d+1)$ vectors ${\bf w} (a\alpha)$ in $\mathbb{C}^{d^2}$, of components $w_{pq}(a\alpha)$, satisfying (\ref{proposition1b}) 
and (\ref{proposition1a}), then we could construct $d(d+1)$ vectors $|a \alpha \rangle$ satisfying (\ref{complete1 def MUBs1}). This can be done by means 
of a diagonalization procedure of the matrices 
			\begin{equation}
M_{a \alpha} = \sum_{p=0}^{d-1} \sum_{q=0}^{d-1} w_{pq}(a \alpha) E_{pq}
			\label{M en termes de E}
			\end{equation}
where $a = 0, 1, \ldots, d$ and $\alpha = 0, 1, \ldots, d-1$. An alternative and more simple way to obtain the $|a \alpha \rangle$ vectors from the ${\bf w} (a\alpha)$ vectors is as follows. Equation (\ref{MUBs en v}) leads to  
  		\begin{equation}
\left\vert \sum_{p=0}^{d-1} \overline{\omega_{p}(a\alpha)} \omega_{p}(b\beta) \right\vert =
\delta_{\alpha ,\beta }\delta_{a,b} + \frac{1}{\sqrt{d}}(1-\delta_{a,b})
			\label{MUBs en v avec racine carree}
  		\end{equation} 
to be compared with (\ref{complete1 def MUBs1}). Then, the $|a \alpha \rangle$ vectors can be constructed once the ${\bf w} (a\alpha)$ vectors are known. The solution, in matrix form, is
			\begin{equation}
|a \alpha \rangle	= \begin{pmatrix} 
\omega_{0}(a\alpha)   \cr 
\omega_{1}(a\alpha)   \cr 
\vdots           \cr
\omega_{d-1}(a\alpha) \cr 
\end{pmatrix}
			\end{equation}
			\begin{equation}	
a = 0, 1, \ldots, d \quad \alpha = 0, 1, \ldots, d-1
			\end{equation}
Therefore, we can construct a complete set $\{ B_a : a = 0, 1, \ldots, d \}$ of $d+1$ MUBs from the knowledge of $d(d+1)$ vectors ${\bf w} (a\alpha)$. Note that, for 
fixed $a$ and $\alpha$, the $|a \alpha \rangle$ vector is an eigenvector of the $M_{a \alpha}$ matrix with the eigenvalue 1. This establishes a link with the above-mentioned diagonalization procedure. 	
			
\section{A parallel problem}			
			
The present work takes its origin in \cite{OA-MRK} where some similar developments were achieved for the search of a SIC POVM. Symmetric informationally complete positive-operator-valued measures play an important role in quantum information. Their existence in arbitrary dimension is still the object of numerous studies (see for instance \cite{AFZ}). 

A SIC POVM in dimension $d$ can be defined as a set of $d^2$ nonnegative operators $P_x = | \Phi_x \rangle \langle \Phi_x |$ acting on $\mathbb{C}^d$ and satisfying 
		  \begin{equation}
\frac{1}{d} \sum_{x = 1}^{d^2} P_x = I 
  		\end{equation}
and		
		  \begin{equation}  		
{\rm Tr} \left( P_x P_y \right) = \frac{d \delta_{x,y} + 1}{d + 1} 		
  		\end{equation}  				
where $I$ is the identity operator. The search for such a SIC POVM amounts to find $d^2$ vectors $| \Phi_x \rangle$ in $C^d$ satisfying 
		  \begin{equation}
\frac{1}{d} \sum_{x = 1}^{d^2} | \Phi_x \rangle \langle \Phi_x | = I
  		\end{equation}
and  		
		  \begin{equation}  		
\vert \langle \Phi_x | \Phi_y \rangle \vert = \sqrt{\frac{d \delta_{x,y} + 1}{d + 1}}  		
  		\end{equation}  		
with $x, y = 1, 2, \ldots, d^2$. The $P_x$ operator can be developed as 
		  \begin{equation}
P_x = \sum_{p=0}^{d-1} \sum_{q=0}^{d-1} v_{pq}(x) E_{pq}
  		\end{equation}
so that the determination of $d^2$ operators $P_x$ (or $d^2$ vectors $| \Phi_x \rangle$) is equivalent 
     to the determination of $d^2$ vectors ${\bf v}(x)$, of components $v_{pq}(x)$, in $C^{d^2}$. In the spirit of the preceding sections, we have the following result. 

{\bf Proposition 3.} {\it For $d \geq 2$, finding a SIC POVM in $\mathbb{C}^d$ (if it exists) is equivalent to finding $d^2$ vectors ${\bf v} (x)$ in $\mathbb{C}^{d^2}$, of components $v_{pq}(x)$ such that
		  \begin{equation}
\frac{1}{d} \sum_{x = 1}^{d^2} v_{pq}(x) = \delta_{p,q}, \quad p,q \in \mathbb{Z}/d\mathbb{Z}
			\label{proposition3a}
  		\end{equation} 		  
		  \begin{equation}
\sum_{p=0}^{d-1} \sum_{q=0}^{d-1} \overline{v_{pq}(x)} v_{pq}(y) = \frac{d \delta_{x,y} + 1}{d + 1}
			\label{proposition3b}
  		\end{equation} 
and  		
			\begin{equation}
v_{pq}(x) = \nu_{p}(x) \overline{\nu_{q}(x)}, \quad p,q \in \mathbb{Z}/d\mathbb{Z}
			\label{proposition3c}
			\end{equation}  		
where $x, y = 1, 2, \ldots, d^2$.}

\section{Concluding remarks}

The equivalence discussed in this work of the two ways of looking at MUBs amounts in some sense to the equivalence between the search for 
equiangular lines in $\mathbb{C}^d$ and for equiangular vectors in $\mathbb{C}^{d^2}$ (cf.~\cite{equiangular1}). Equiangular lines in $\mathbb{C}^d$ correspond to 
			\begin{eqnarray}
| \langle a\alpha | b\beta \rangle | = \frac{1}{\sqrt{d}} \ {\rm for} \ a \not= b 								
			\label{equi lines} 
			\end{eqnarray}
while equiangular vectors in $\mathbb{C}^{d^2}$ correspond to 
			\begin{eqnarray}
{\bf w} (a\alpha) \cdot {\bf w}(b\beta) = \frac{1}{d} \ {\rm for} \ a \not= b
			\label{equi vectors} 
			\end{eqnarray}
where the ${\bf w} (a\alpha) \cdot {\bf w}(b\beta)$ inner product in $\mathbb{C}^{d^2}$ is defined as
			\begin{eqnarray}
{\bf w} (a\alpha) \cdot {\bf w} (b\beta) = \sum_{p=0}^{d-1} \sum_{q=0}^{d-1} \overline{w_{pq}(a\alpha)} w_{pq}(b\beta) 
			\label{ps en w} 
			\end{eqnarray}
Observe that the modulus disappears and the $1/{\sqrt{d}}$ factor is replaced by ${1}/{d}$ when passing from (\ref{equi lines}) to (\ref{equi vectors}). It was questioned in~\cite{Kib13} if the equiangular vectors approach can shed light on the still unsolved question to know if one can find d+1 MUBs when $d$ is not a (strictly positive) power of a prime integer. In the case where $d$ is not a power of a prime, the impossibility of finding $d(d + 1)$ vectors ${\bf w} (a\alpha)$ or $d(d + 1)$
matrices $M_{a\alpha}$ satisfying the conditions in Propositions 1 and 2 would mean that $d+1$ MUBs do not exist in $\mathbb{C}^d$. However, it is hard to know if one approach is better than the other. It is the hope of the author that the equiangular vectors approach be tested in the $d = 6$ case for which one knows only three MUBs instead of $d+1 = 7$ in spite of numerous numerical studies (see~\cite{dim61,dim62,dim63} and references therein for an extensive list of related works). 

Similar remarks apply to SIC POVMs. The existence problem of SIC POVMs in arbitrary dimension is still unsolved although SIC POVMs have been constructed in every dimension $d \leq 67$ (see \cite{AFZ} and references therein). For SIC POVMs, the equiangular lines in $\mathbb{C}^d$ correspond to 
			\begin{eqnarray}
\vert \langle \Phi_x | \Phi_y \rangle \vert = \frac{1}{\sqrt{{d + 1}}} \ {\rm for} \ x \not= y 								
			\label{equi lines SIC} 
			\end{eqnarray}
and the equiangular vectors in $\mathbb{C}^{d^2}$ to 
			\begin{eqnarray}
{\bf v} (x) \cdot {\bf v} (y) = \frac{1}{d+1} \ {\rm for} \ x \not= y
			\label{equi vectors SIC} 
			\end{eqnarray}
where the ${\bf v} (x) \cdot {\bf v} (y)$ inner product in $\mathbb{C}^{d^2}$ is defined as
			\begin{eqnarray}
{\bf v} (x) \cdot {\bf v} (y) = \sum_{p=0}^{d-1} \sum_{q=0}^{d-1} \overline{v_{pq}(x)} v_{pq}(y) 
			\label{ps en w SIC} 
			\end{eqnarray}

The parallel between MUBs and SIC POVM characterized by the couples of equations (\ref{equi lines})-(\ref{equi lines SIC}),  
(\ref{equi vectors})-(\ref{equi vectors SIC}) and (\ref{ps en w})-(\ref{ps en w SIC}) should be noted. These matters shall 
be the subject of a future work. 

\begin{acknowledgements}
The material contained in the present note was planned to be presented at the eleventh edition of the workshop {\it Analytic and Algebraic 
Methods in Physics} (AAMP~XI). Unfortunately, the author was unable to participate in AAMP~XI. He is greatly indebted to Miloslav Znojil for 
suggesting him to submit this work to Acta Polytechnica. 
\end{acknowledgements}

\bibliographystyle{actapoly}
\bibliography{actapoly_biblio}

\end{document}